\documentclass{iaus}
\usepackage{graphicx}

\title[Colliding winds in $\eta$~Carinae] 
{3-D SPH simulations of colliding winds in {\boldmath $\eta$}~Carinae}

\author[A. T. Okazaki, S. P. Owocki, C. M. P. Russell, \&
M. F. Corcoran]   
{Atsuo T. Okazaki$^1$, Stanley P. Owocki$^2$, Christopher M. P. Russell$^3$,
\and Michael F. Corcoran$^4$}

\affiliation{$^1$Faculty of Engineering, Hokkai-Gakuen University, \\
Toyohira-ku, Sapporo 062-8605, Japan \\ 
email: {\tt okazaki@elsa.hokkai-s-u.ac.jp} \\[\affilskip]
$^2$Bartol Research Institute, University of Delaware, \\
Newark, 19716 DE, USA \\
email: {\tt owocki@bartol.udel.edu} \\[\affilskip]
$^3$Department of Physics and Astronomy, University of Delaware, \\
Newark, 19716 DE, USA \\
email: {\tt crussell@udel.edu} \\[\affilskip]
$^4$Universities Space Research Association, Goddard Space Flight Center, \\
Greenbelt, MD 20771, USA \\ 
email: {\tt corcoran@milkyway.gsfc.nasa.gov}}

\pubyear{2008}
\volume{250}  
\pagerange{1--6}
\jname{Massive Stars as Cosmic Engines}
\editors{F. Bresolin, P.A. Crowther \& J. Puls, eds.}
\begin{document}

\maketitle

\begin{abstract}
We study colliding winds in the superluminous binary $\eta$ Carinae by performing 
three-dimensional, Smoothed Particle Hydrodynamics (SPH) simulations.
For simplicity, we assume both winds to be isothermal. We also assume 
that wind particles coast without any net external forces.
We find that the lower density, faster wind from the secondary carves out 
a spiral cavity in the higher density, slower wind from the primary.
Because of the phase-dependent orbital motion, the cavity is very thin 
on the periastron side,
whereas it occupies a large volume on the apastron side.
The model X-ray light curve using the simulated density structure fits very well
with the observed light curve for a viewing angle of
$i = 54^\circ$ and $\phi = 36^\circ$,
where $i$ is the inclination angle and $\phi$ is the azimuth 
from apastron.
\keywords{stars: luminous blue variable, winds -- hydrodynamics -- methods: numerical 
-- binaries: general}
\end{abstract}

\firstsection 
\section{Introduction}

$\eta$~Carinae is one of the most luminous and massive stars in the Galaxy.
It has exhibited a series of mass ejection episodes, the most notable of which 
was the Great eruption in the 1840's when the star ejected mass of 
$\sim 10\,M_{\odot}$, from which the Homunculus nebula was formed. 
Its current mass and luminosity are 
$M \sim 10^{2}\,M_{\odot}$ and $L \sim 5 \times 10^{6}\,L_{\odot}$,
respectively.

The spectrum of $\eta$ Car is rich in emission lines and has no photospheric lines.
\cite[Damineli (1996)]{dam96} first noticed a 5.5~yr periodicity in the variability of
the He\,I\,10830\AA\ line. Later, other optical lines were also found to
show variations with the same periodicity. 
In the X-ray band, the flux exhibits particularly interesting, periodic variability: 
After a gradual increase toward periastron, the X-ray flux suddenly drops to a minimum, 
which lasts for about three months. It then recovers to a level 
slightly higher than that at apastron 
(\cite[Ishibashi et al. 1999]{osh99}; \cite[Corcoran 2005]{cor05};
see also Fig.~\ref{fig:okazaki-fig3}).
All these variations are
consistent with the system being a long-period ($P_{\rm orb} = 2,024$ days), 
highly eccentric ($e \sim 0.9$) binary. The X-ray emission is considered to 
arise from the wind collision region.

Although there is mounting evidence that $\eta$~Car is a supermassive binary,
it is very hard to directly observe the stars.
They are buried deep inside dense winds, which are further engulfed by 
the optically thick, Homunculus nebula. As a result, even the viewing angle is not
well constrained. It is therefore important to construct a 3-D
dynamical model, on the basis of which the observed features are interpreted.

In this paper, we give a brief summary of the results from 3-D numerical simulations
of colliding winds in $\eta$ Car.
Detailed results will be published elsewhere (\cite[Okazaki et al. 2008]{oka08}).

\section{Numerical Model}

Simulations presented here were performed with a 3-D Smoothed
Particle Hydrodynamics (SPH) code. The code is based on a version 
originally developed by Benz (\cite[Benz 1990]{ben90a};
\cite[Benz et al. 1990]{ben90b}) and then 
by Bate and his collaborators (\cite[Bate, Bonnell \& Price 1995]{bat95}). 
It uses the variable smoothing length, and the SPH equations with
the standard cubic-spline kernel are integrated with individual time steps for
each particle. In our code, the winds are modeled by
an ensemble of gas particles, which are continuously ejected with
a given outward velocity at a radius just outside each star.
The artificial viscosity parameters adopted are $\alpha_{\rm SPH}=1$ and 
$\beta_{\rm SPH}=2$.

For simplicity, we take both winds to be isothermal and coasting
without any net external forces, assuming in effect that 
gravitational forces are effectively canceled by radiative driving terms. 
We set the binary orbit 
on the $x$-$y$ plane and the major axis of the orbit  
along the $x$-axis (the apastron is in the $+x$-direction). 
The outer simulation boundary is set at either $r=10.5a$ 
or $r = 105a$ from
the centre of mass of the system, where $a$ is the semi-major axis of 
the binary orbit. Particles crossing this boundary are removed
from the simulation. 
In the following, $t=0$ (Phase 0) corresponds to the periastron passage.

Table~\ref{tbl:params} summarizes the stellar, wind, and orbital parameters 
adopted in our simulations.
With these parameters, the ratio $\eta$ of the momentum fluxes of the winds
from $\eta$ Car A and B is $\eta \sim 4.2$.
The parameters adopted here are
consistent with those derived from observations (\cite[Corcoran et al. 2001]{cor01};
\cite[Hillier et al. 2001]{hil01}),
except for the wind temperature of $\eta$ Car A. As mentioned above, we take the same 
temperature for both winds for simplicity. Note that the effect of wind temperature on 
the dynamics of high-velocity wind collision is negligible.

\begin{table}
  \begin{center}
  \caption{Stellar, wind, and orbital parameters}
  \label{tbl:params}
  \begin{tabular}{lcc}
  \hline
  \multicolumn{1}{c}{Parameters}
  & $\eta$ Car A & $\eta$ Car B \\
  \hline
  Mass ($M_{\odot}$) & 90 & 30 \\
  Radius ($R_{\odot}$) & 90 & 30 \\
  Mass loss rate ($M_{\odot}\,{\rm yr}^{-1}$) & $2.5 \times 10^{-4}$ & $10^{-5}$ \\
  Wind velocity (${\rm km\,s}^{-1}$) & 500 & 3,000 \\
  Wind temperature (K) & $3.5 \times 10^{4}$ & $3.5 \times 10^{4}$ \\
  \hline
  Orbital period $P_{\rm orb}$ (d) & \multicolumn{2}{c}{2,024} \\
  Orbital eccentricity $e$ & \multicolumn{2}{c}{0.9} \\
  Semi-major axis $a$ (cm) & \multicolumn{2}{c}{$2.3 \times 10^{14}$} \\
  \hline
  \end{tabular}
  \end{center}
\end{table}

\section{Structure and evolution of colliding winds}

\begin{figure}[!t]
\begin{center}
 \includegraphics[width=11cm]{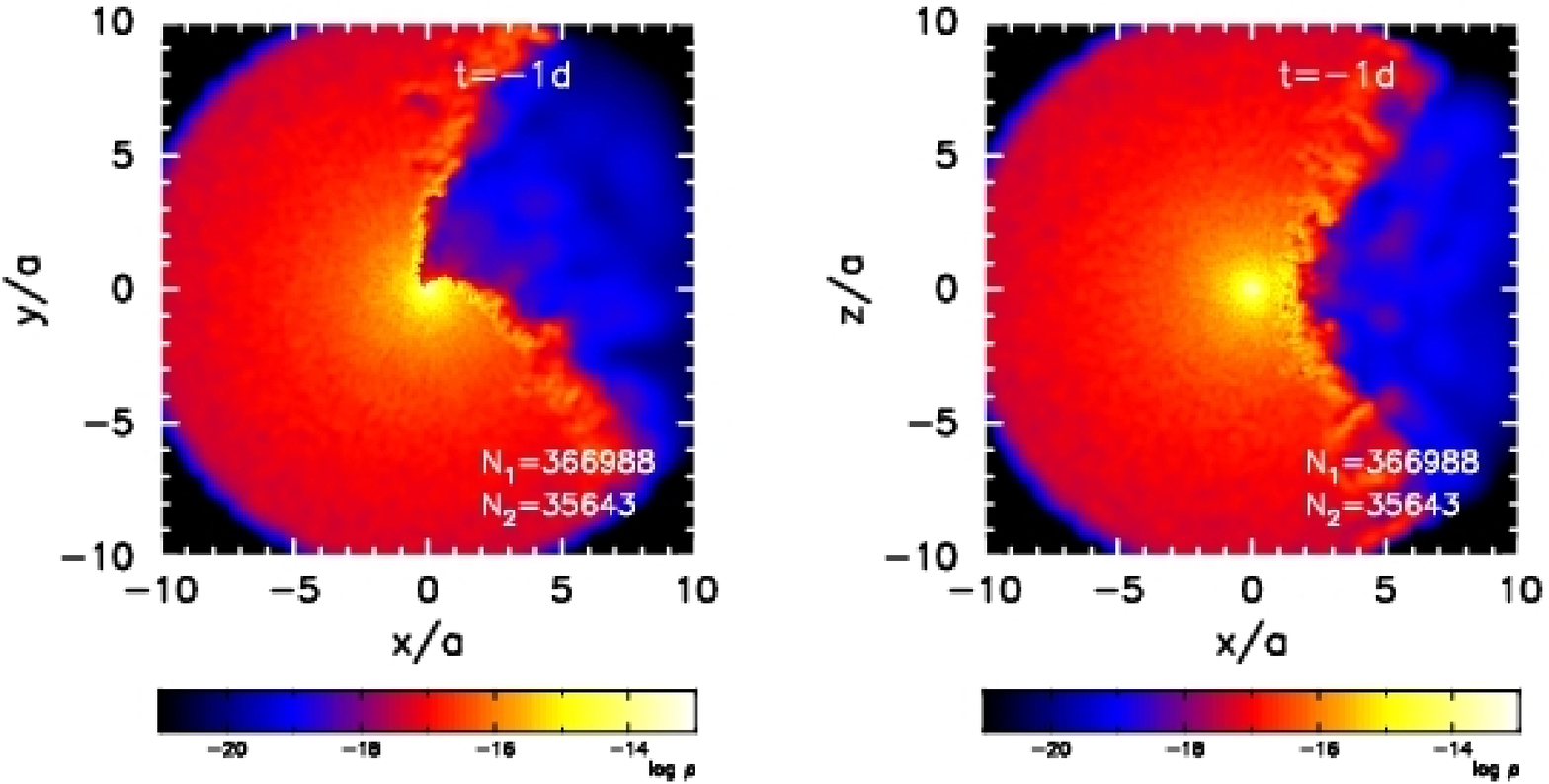} \\
 \includegraphics[width=11cm]{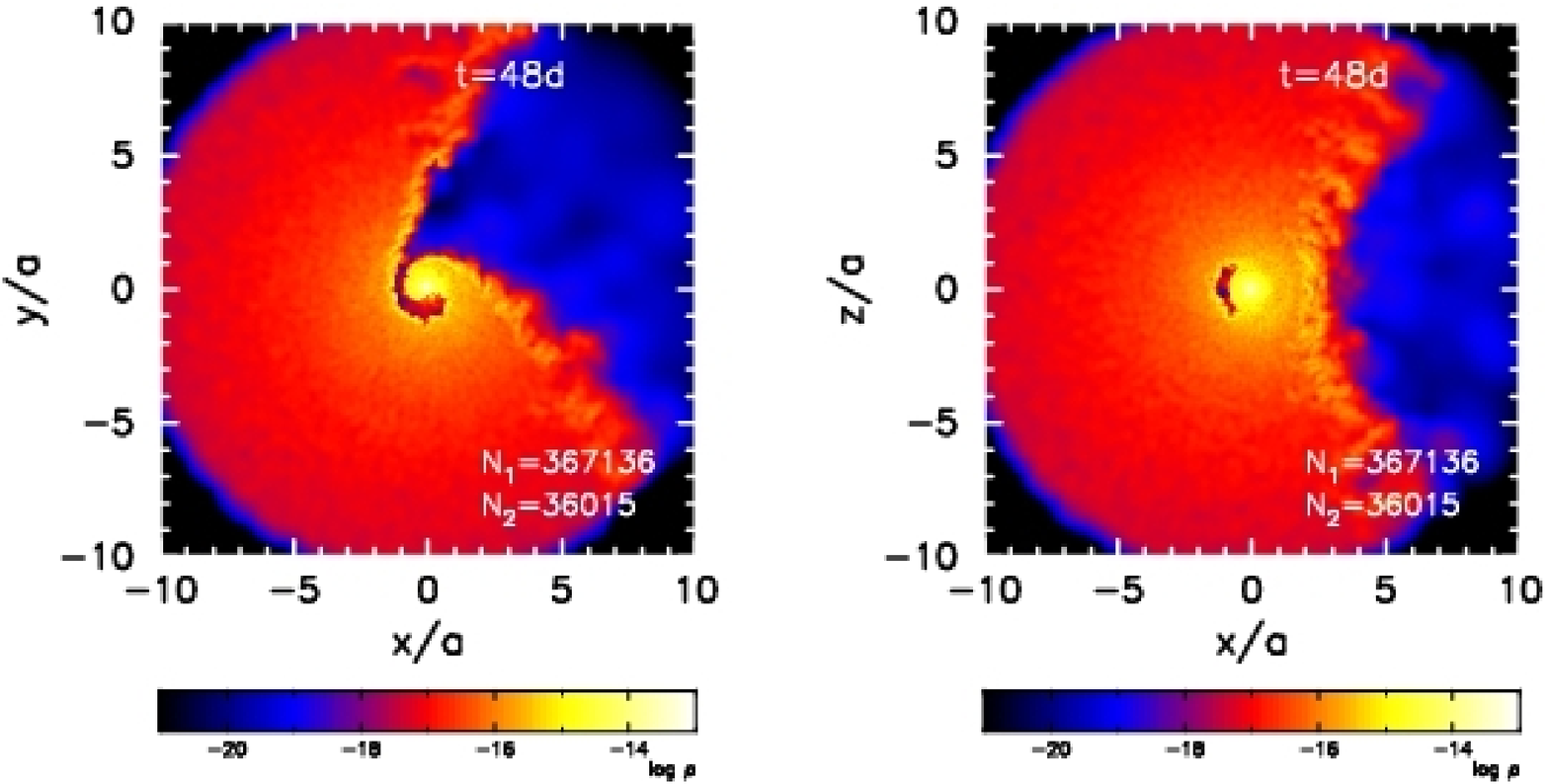} \\
 \includegraphics[width=11cm]{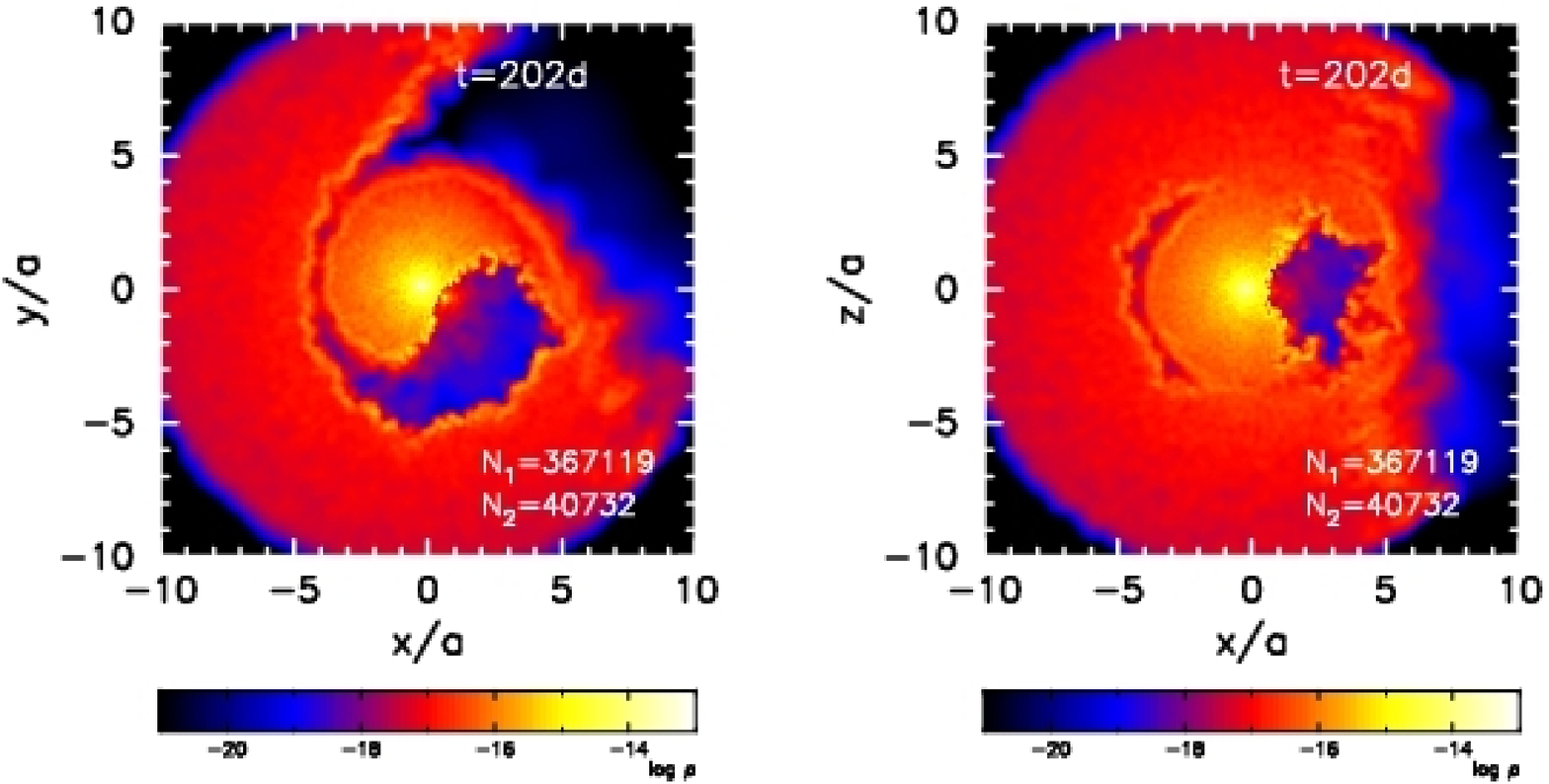} 
 \caption{Wind collision interface geometry at $t \sim 0$ (top), $t \sim 50\,{\rm d}$
    (middle), and $t \sim 200\,{\rm d}$ (bottom) for $r \le 10a$. 
    In each panel, the greyscale plot shows the density in the orbital plane (left) 
and the plane which is perpendicular to the orbital plane and through 
the major axis of the orbit (right),
on a logarithmic scale with cgs units.
The dark spot near the origin represents the primary, while
the small dark spot close to the apex of the lower density wind represents 
the secondary.
Annotations in each panel give the time (in days) from periastron passage and the
numbers of particles in the primary wind, $N_{1}$, and in the secondary wind, $N_{2}$.}
 \label{fig:okazaki-fig1}
\end{center}
\end{figure}

Figure~\ref{fig:okazaki-fig1} shows the wind collision interface geometry 
at $t \sim 0$ (at periastron; top panel), $t \sim 50\,{\rm d}$ 
(during the X-ray minimum; middle panel), and 
$t \sim 200\,{\rm d}$ (after the X-ray minimum; bottom panel) in 
a simulation covering $r \le 10.5a$.
In each panel, the greyscale plot shows the density
in the orbital plane (left panel)
and the plane perpendicular to the orbital plane and through 
the major axis of the orbit (right panel),
on a logarithmic scale with cgs units.
The dark spot near the origin represents the primary ($\eta$ Car A), while
the small dark spot close to the apex of the lower density wind represents 
the secondary ($\eta$ Car B).

Although the wind collision interface in the simulation exhibits variations from 
instabilities, its global shape is easily traced and illustrates 
how the lower density,
faster wind from the secondary makes a cavity in the higher density, slower wind
from the primary.
As expected, the shape of the collision interface around apastron, 
where the orbital speed 
of the secondary is only $\sim 20\,{\rm km\,s}^{-1}$ with respect to the primary,
is in agreement with the analytical one (e.g., \cite[Antokhin et al. 2004]{ant04}). 
As the secondary approaches the periastron, 
the interface begins to bend, and at phases around periastron, 
where the orbital speed of the secondary
is $\sim 360{\rm km\,s}^{-1}$ with respect to the primary, 
the lower density wind from the secondary makes
a thin layer of cavity along the orbit. Then, the thickness of the cavity increases 
as the secondary moves away from periastron.

\begin{figure}[!t]
\begin{center}
 \includegraphics[width=11cm]{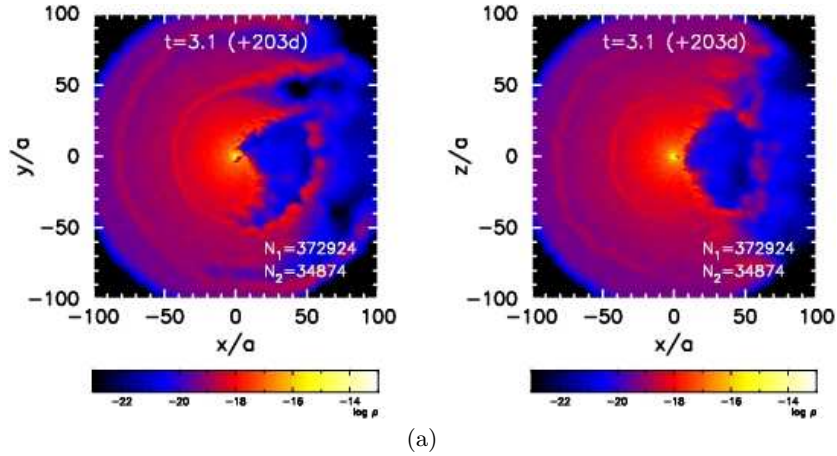} \\
 (a) \\
 \includegraphics[width=8cm]{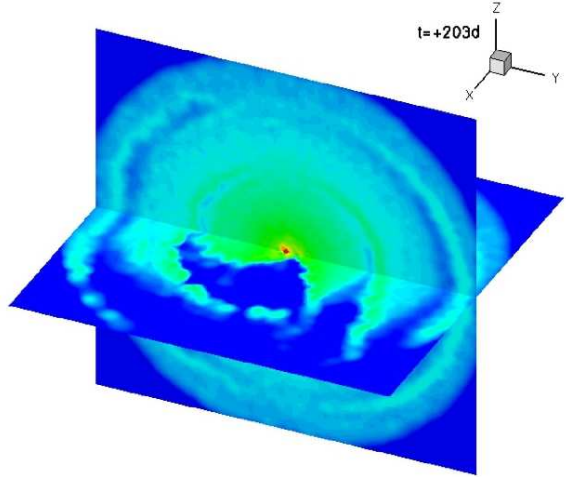} \\
 (b)
 \caption{Wind collision interface geometry at $t \sim 200\,{\rm d}$ 
 within $r = 100a$: (a) 2-D density maps in the orbital plane (left)
and the plane perpendicular to the orbital plane and through 
the major axis of the orbit (right) and (b) the 3-D plot of the 
logarithmic density.}
 \label{fig:okazaki-fig2}
\end{center}
\end{figure}

In order to study the wind collision interface geometry on a larger scale,
we have performed a simulation covering $r \le 105a$. 
Figure~\ref{fig:okazaki-fig2}(a) shows the 2-D density maps in the orbital plane 
and the plane perpendicular to it at the same phase (Phase 0.1) as that 
of the bottom panel of Fig.~\ref{fig:okazaki-fig1}. The 3-D interface geometry 
is shown in Fig.~\ref{fig:okazaki-fig2}(b) by the logarithmic density plot.
From Fig.~\ref{fig:okazaki-fig2}, we note that the lower density wind from the secondary 
carves out a large-scale, spiral cavity in 
the higher density wind from the primary. The shape of the cavity is very asymmetric. 
It is just a thin layer on the periastron side with respect to the primary, 
whereas it occupies a large volume on the apastron side.

\begin{figure}[!t]
\begin{center}
 \includegraphics[width=8cm]{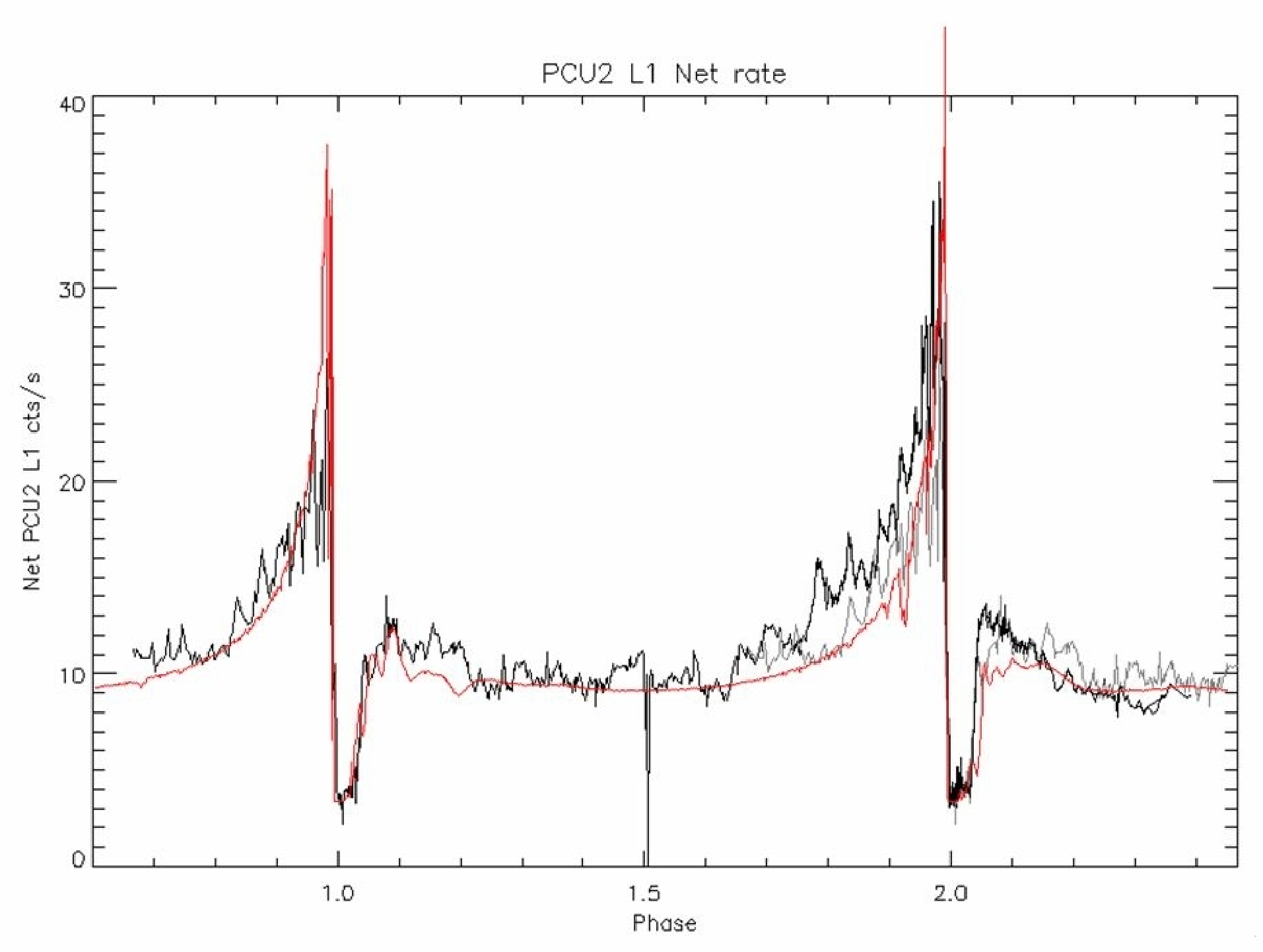} 
 \caption{Comparison of RXTE data (black and gray lines; 
 see \cite[Corcoran 2005]{cor05}) 
 with the closest match model light curve (less-jaggy gray line). 
 The light gray line shows the first cycle of RXTE data shifted by one period.
 Taken from \cite[Russell et al. (2008)]{rus08}.}
 \label{fig:okazaki-fig3}
\end{center}
\end{figure}

It is interesting to study whether the model presented here can explain the variability 
in the RXTE X-ray light curve.
Using the density distribution in our $r=10a$ simulation, 
\cite[Russell et al. (2008)]{rus08} modeled the 
X-ray light curve and compared it with the observed light curve. 
Assuming that the X-ray emission occurs at the head 
of the wind-wind interaction cone located at $D/(1+\sqrt{\eta})$
from the secondary, where $D$ is the binary separation, 
and varies in intensity with $D^{-1}$ at any given orbital phase, 
they generated trial X-ray light curves by computing the phase variation of 
absorption to observers at various assumed lines of sight.
They found that the RXTE X-ray light curve is very well fit with an optimal viewing angle
of approximately 54 degrees of inclination and 36 degrees from apastron 
in the prograde direction
(see Fig.~\ref{fig:okazaki-fig3} from \cite[Russell et al. (2008)]{rus08}
for the comparison between the model and 
observed X-ray light curves).
The excellent fit seen in Fig.~\ref{fig:okazaki-fig3} confirms that the current model
basically gives a correct picture of the wind-wind collision interaction in
$\eta$~Carrinae.

\section{Conclusions}
We have studied the wind collision interaction in the supermassive binary
$\eta$ Carinae, carrying out
3-D SPH simulations. The results from 
simulations have clarified how the lower density, faster wind from the secondary
($\eta$ Car B) carves out a cavity in 
the higher density, slower wind from the primary ($\eta$ Car A).
With an optimal viewing angle of $i = 54^\circ$ and $\phi = 36^\circ$,
where $i$ is the inclination angle and $\phi$ is the azimuth measured 
from apastron in the prograde direction, the model gives an excellent fit
with the RXTE X-ray light curve
(\cite[Russell et al. 2008]{rus08}).

\begin{acknowledgements}
A.T.O. thanks Japan Society for the Promotion of Science for
the financial support via Grant-in-Aid for Scientific Research
(16540218).
SPH simulations were performed on HITACHI SR11000 at Hokkaido
University Information Initiative Center.
\end{acknowledgements}

\begin{discussion}

\discuss{Davidson}{Four essential points.
\begin{enumerate}
\item One cannot derive useful orbit parameters for this object from
Doppler velocities, because \emph{every} available spectral feature
evolves in a complex way. Eccentricity 0.9 is possible but not established.
\item Several years ago, Kazunori Ishibashi proposed an orbit orientation
based on the X-rays. It roughly matched the parameters you adopted.
\item $\eta$~Car's spectroscopic events are \emph{not} primarily eclipses,
not even eclipses by the wind. He~II, He~I, near IR, photometry, and X-ray flares
all show that something \emph{far} more interesting is involved.
See J. Martin et al. 2006, \textit{ApJ} and references therein.
\item The wind-wind collision region is at low latitudes
but nearly all of the primary wind is polar (except during an event!).
\end{enumerate}}

\discuss{Kudritzki}{The X-ray dip is not an eclipse in the classical stellar sense,
but rather a \lq \lq wind eclipse'' or an interval when the X-ray emission 
from the wind-wind collision is embedded in the dense wind from the primary.}

\discuss{Okazaki}{Studying the effect of such an asymmetry in the primary wind
on the interaction geometry is interesting. It is easy to implement 
the wind asymmetry in my code.}

\end{discussion}


\begin{thebibliography}{99}

\bibitem[Antokhin et al. (2004)]{ant04}
  {Antokhin, I. I., Owocki, S. P., \& Brown, J. C.} 2004, ApJ, 611, 434

\bibitem[Bate et al. (1995)]{bat95}
  {Bate, M.R., Bonnell, I.A., \& Price, N.M.} 1995, \textit{MNRAS}, 285, 33

\bibitem[Benz (1990)]{ben90a}
  {Benz, W.} 1990, In: J.R. Buchler (ed.), 
  \textit{The Numerical Modelling of Nonlinear Stellar Pulsations}
  (Dordrecht: Kluwer), p.\,269

\bibitem[Benz et al. (1990)]{ben90b}
  {Benz, W., Bowers, R.L., Cameron, A.G.W., \& Press, W.H.} 1990, 
  \textit{ApJ}, 348, 647

\bibitem[Corcoran et al. (2001)]{cor01}
  {Corcoran, M.F., Ishibashi, K., Swank, J.H., \& Petre, R.} 2001, 
  \textit{ApJ}, 547, 1034

\bibitem[Corcoran (2005)]{cor05}
  {Corcoran, M.F.} 2005, \textit{AJ}, 129, 2018

\bibitem[Damineli (1996)]{dam96}
  {Damineli, A.} 1996, \textit{ApJ}, 460, L49

\bibitem[Hillier et al. (2001)]{hil01}
  {Hillier, D.J., Davidson, K., Ishibashi, K., \& Gull, T.} 2001,
  \textit{ApJ}, 553, 837

\bibitem[Ishibashi et al. (1999)]{ish99}
  {Ishibashi, K., Corcoran, M.F., Davidson, K., Swank, J.H., 
  Petre, R., Drake, S.A., Damineli, A., \& White, S.} 1999, \textit{ApJ}, 524, 983

\bibitem[Okazaki et al. (2002)]{oka02}
  {Okazaki, A.T., Bate, M.R., Ogilvie, G.I, \& Pringle, J.E.} 2002, 
  \textit{MNRAS}, 337, 967

\bibitem[Okazaki et al. (2008)]{oka08}
  {Okazaki, A.T., Owocki, S.P., Russell, C.M.P., \& Corcoran, M.F.} 2008, 
  in preparation

\bibitem[Russell et al. (2008)]{rus08}
  {Russell, C.M.P., Owocki, S.P., \& Okazaki, A.T.} 2008, these proceedings

\end{thebibliography}
\end{document}